\documentclass[twocolumn,preprintnumbers,amsmath,amssymb]{revtex4}

\usepackage{graphicx}
\usepackage{dcolumn}
\usepackage{bm}

\begin{document}

\title{Oscillations of the large-scale circulation in turbulent Rayleigh-B\'{e}nard convection: the off-center mode and its relationship with the torsional mode}
\author{Quan ZHOU, Heng-Dong XI, Sheng-Qi ZHOU, Chao SUN and Ke-Qing XIA}
\address{Department of Physics, The Chinese University of Hong Kong,
Shatin, Hong Kong, China}
\date{\today}

\begin{abstract}

We report an experimental study of the large-scale circulation
(LSC) in a turbulent Rayleigh-B\'{e}nard convection cell with
aspect ratio unity. The temperature-extremum-extraction (TEE)
method for obtaining the dynamic information of the LSC is
presented. With this method, the azimuthal angular positions of
the hot ascending and cold descending flows along the sidewall are
identified from the measured instantaneous azimuthal temperature
profile. The motion of the LSC is then decomposed into two
different modes based on these two angles: the azimuthal mode and
the translational or off-center mode that is perpendicular to the
vertical circulation plane of the LSC. Comparing to the previous
sinusoidal-fitting (SF) method, it is found that both the TEE and
the SF methods give the same information about the azimuthal
motion of the LSC, but the TEE method in addition can provide
information about the off-center motion of the LSC. The off-center
motion is found to oscillate time-periodically around the cell's
central vertical axis with an amplitude being nearly independent
of the turbulent intensity. It is further found that the azimuthal
angular positions of the hot ascending flow near the bottom plate
and the cold descending flow near the top plate oscillate
periodically out of phase by $\pi$, leading to the torsional mode
of the LSC. These oscillations are then propagated vertically
along the sidewall by the hot ascending and cold descending
fluids. When they reach the mid-height plane, the azimuthal
positions of the hottest and coldest fluids again oscillate out of
phase by $\pi$. It is this out-of-phase horizontal positional
oscillation of the hottest and coldest fluids at the same
horizontal plane that produces the off-center oscillation of the
LSC. A direct velocity measurement further confirms the existence
of the bulk off-center mode of the flow field near cell center.

The paper has been submitted to \emph{\textbf{J. Fluid Mech.}}

\end{abstract}

\maketitle

\section{Introduction}

The convection phenomenon is ubiquitous in nature and in many
engineering applications. A simple but paradigmatic model that has
been widely used to study the convection phenomenon for more than
a century is the turbulent Rayleigh-B\'{e}nard (RB) convection,
which is a fluid layer heated from below and cooled on the top
(for recent reviews, see \cite{siggia1994arfm, agl2008rmp}). The
dynamics of the RB system is determined by its geometry and two
dimensionless control parameters: The Rayleigh number $Ra=\beta
gH^3\Delta T/\nu\kappa$ and the Prandtl number $Pr=\nu/\kappa$,
where $g$ is the gravitational acceleration, $H$ the height of the
cell, $\Delta T$ the temperature difference across the cell, and
$\beta$, $\nu$, and $\kappa$, respectively, the thermal expansion
coefficient, the kinematic viscosity, and the thermal diffusivity
of the fluid. The geometry of the system is described by its
symmetry and the aspect ratio of the convection cell $\Gamma$. For
a cylindrical cell, $\Gamma=D/H$ with $D$ being the cell's
diameter.

A prominent feature of turbulent RB system is the presence of the
large-scale circulation (LSC), which is self-organized from
thermal plumes that erupt from the top and bottom thermal boundary
layers \cite{xi2004jfm}. We call it ``large-scale" because it is a
single cellular structure that spans the height of the cell, at
least in cells with aspect ratios close to one. In an axial
symmetric configuration, the LSC is found to exhibit stochastic
azimuthal meandering and small-probability events of cessations
and reversals (see \cite{agl2008rmp} and references therein). In
addition, the horizontal motions of the LSC near the top and
bottom plates are found to oscillate periodically out of phase by
$\pi$. This twisting oscillation of the LSC's vertical circulation
plane is referred to as the torsional mode of the LSC, which was
first found by Funfschilling $\&$ Ahlers
\cite{funfschilling2004prl}.

In addition to the torsional oscillation of the LSC, there also
exists a well-defined low-frequency oscillation in both the
temperature and velocity fields, which has been long observed in
turbulent RB experiments using various fluids (see, e.g.,
\cite{castaing1989jfm, takeshita1996prl, ashkenazi1999prl,
qiu2001prl, qiu2002pre, lam2002pre}). As oscillation is a common
phenomenon in closed flow systems, understanding the nature and
the origin of the low-frequency oscillation in the RB system
should be of general interest. To understand this oscillation,
Villermaux \cite{villermaux1995prl} proposed the coupled boundary
layer model and the two main predictions of the model are: (1)
Plumes are emitted periodically and (2) hot and cold plumes are
emitted alternatively. The two predictions are not independent of
each other. Physically speaking, thermal plumes can emit
periodically but not necessarily alternatively. However, they
cannot emit both alternatively and nonperiodically. This is
because the alternative emission of hot and cold plumes implies a
triggering mechanism of plume generation, i.e., hot plumes cause
the emission of cold ones and vice versa. Therefore, the emission
of one type of plumes should wait for the arrival of the other
type. Since there exists a typical timescale for thermal plumes to
travel from one plate to the other, the alternative emission
process should quickly lead to a synchronized and periodic
emission of hot and cold plumes from the bottom and top plates
respectively. Some experiments based on single-point or
two-dimensional (2D) measurements of the temperature and velocity
fields appear to support Villermaux's model
\cite{ciliberto1996pre, qiu2001prl, qiu2002pre, sun2005pre,
tsuji2005prl}. Some other works, however, pointed out that
periodic plume emission is not necessary for the periodicity of
the system \cite{funfschilling2004prl, resagk2006pof}. Recently,
Ahlers \emph{et al.} \cite{agl2008rmp} conjectured that the
low-frequency oscillation of the system presumably is due to the
torsional oscillation of the LSC. However, how the torsional mode
generates the temperature and velocity oscillations at the
mid-height plane of the cell is unclear, since, without the
off-center oscillation, a simple twisting oscillation near the top
and bottom plates would cancel out at the mid-height plane due to
symmetry.

In a recent experimental study of the three-dimensional (3D)
structure of temperature oscillations, Xi \emph{et al.}
\cite{xi2008prl} have presented convincing evidences that hot and
cold thermal plumes are emitted neither periodically nor
alternatively, but continuously and randomly, from the top and
bottom plates. Thus invalidate the two predictions of the
Villermux's model. They further showed that the low-frequency
oscillation at the cell's mid-height plane is due to the
off-center oscillation of the LSC. It should be mentioned that a
previous single-point velocity measurement has shown that at the
cell center the strongest velocity oscillation is along the
direction perpendicular to the LSC plane and that the strength of
this oscillation decays away from the cell center towards the
plates \cite{qiu2004pof}. More recently, a study of the 3D spatial
structure of the velocity field has shown that the velocities
along the axis perpendicular to the LSC plane and at the cell's
mid-height plane correlate strongly with each other and have a
common phase across the cell's entire diameter \cite{sun2005pre}.
Both these results imply the existence of the off-center motion.
However, the nature of this motion and its relationship with the
torsional oscillation of the LSC have not been revealed, which are
among the objectives of this paper.

The rest of the paper is organized as follows. We describe the
experimental setup and conditions in Section 2.1. In Section 2.2,
we describe in detail a method for extracting the azimuthal
angular positions of the hottest and coldest fluids along the
sidewall and the LSC's central line from the measured
instantaneous azimuthal temperature profile and provide a
validation of this method. Comparisons with the previous
sinusoidal-fitting method will also be made. The experiment
results are presented and analyzed in Section 3, which is divided
into three parts. In Section 3.1 we present a detailed study of
the off-center oscillation of the LSC. Section 3.2 discusses the
relationship between the off-center mode and the torsional mode of
the LSC and Section 3.3 presents results from a direct velocity
measurement of the bulk off-center mode of the flow field. We
summarize our findings and conclude in Section 4.

\section{Experimental setup, methods and parameters}
\label{sec:multi-thermal-probe-mothod}

\subsection{The convection cell and experiment conditions}

\begin{figure}
\center
\resizebox{1\columnwidth}{!}{%
  \includegraphics{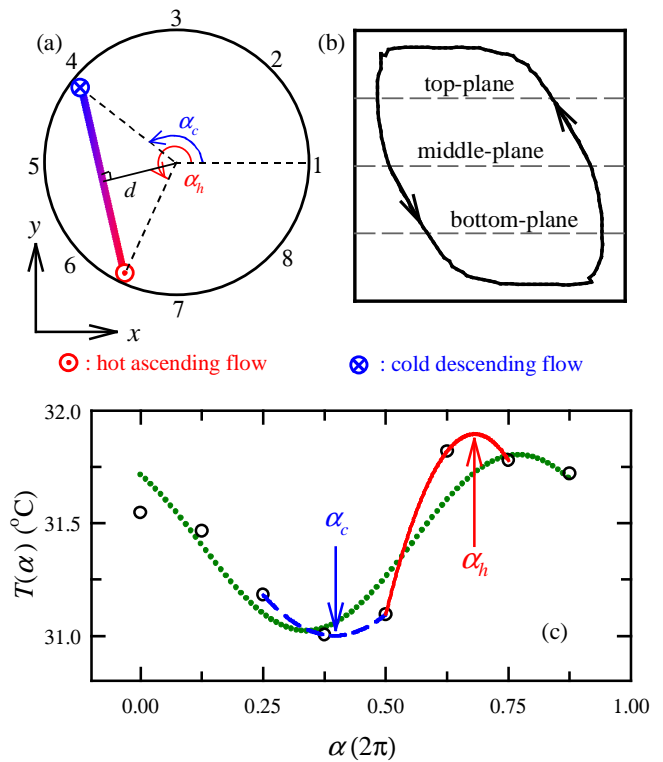}
}\caption{(a) Top-view of the convection cell and coordinate
system used. Also shown is an actual instantaneous azimuthal
positions $\alpha_{h}$ and $\alpha_{c}$ of the hot ascending and
cold descending flows, corresponding to the temperature profile
$T(\alpha)$ shown in (c). The $z$-axis (not shown in the figure)
is along the cell's central vertical axis. The numbers
$1,2,\ldots,8$ show the locations of the sidewall thermistors. (b)
Side-view of the cell with measured circulation path of the LSC
reproduced from Ref. \cite{sunxia2005pre}. The three dashed lines
show the top, middle and bottom planes where the thermistors are
placed. (c) An example of an actual instantaneous temperature
profile (circles) measured by the 8 thermistors at mid-height
plane at $Ra=5.5\times10^9$. The solid line represents a quadratic
fit to the highest temperature reading and the two temperature
readings adjacent to it. The dashed line represents a quadratic
fit to the lowest temperature reading and the two temperature
readings adjacent to it. The two vertical arrows indicate the
fitted peak positions of $\alpha_{h}$ and $\alpha_{c}$,
respectively. The dotted line shows a sinusoidal fit to the
profile. It is seen from the figure that, in this case, the
sinusoidal fit to the profile shifts $\alpha_{h}$ to the right
(anticlockwise in (a)) and shifts $\alpha_{c}$ to the left
(clockwise in (a)). }\label{fig:fig1}
\end{figure}

The experiment was carried out in a cylindrical cell with its top
and bottom plates made of 1.5-cm-thick copper, the sidewall of a
5-mm-thick Plexiglas tube and water as convection fluid
\cite{xi2008pof}. The inner diameter $D$ and the height $H$ of the
cell are both 19.0 cm and hence its aspect ratio $\Gamma$ is
unity. Twenty-four thermistors (Omega, 44031) with a diameter of
2.5 mm and an accuracy of $0.01^{\circ}$ were placed in blind
holes drilled horizontally from the outside into the sidewall with
a distance of 0.7 mm from the fluid-contact surface. These
thermistors are distributed in three horizontal rows at distances
$H/4$, $H/2$, and $3H/4$ from the bottom plate, which are denoted
as the bottom, middle and top planes (the three dashed lines in
figure \ref{fig:fig1}(b)), respectively, and in eight vertical
columns equally spaced azimuthally around the cylinder (figure
\ref{fig:fig1}(a)). A multichannel multimeter was used to record
the resistances of the 24 thermistors at a sampling rate of 0.29
Hz, which are converted into temperatures using calibration
curves. Then the azimuthal temperature profile $T(\alpha)$ at the
three heights can be obtained, where $\alpha$ is the azimuthal
angle referenced to location 1. During the measurements, the mean
temperature $T_0$ of the bulk fluid was kept at $31^{\circ}$C and
hence $Pr=5.3$. The measurements covered six values of $Ra$,
ranging from $9.0\times10^8$ to $6.0\times10^9$, and lasted 70 to
750 hours.

In Ref. \cite{xi2008prl} the cell was tilted by $2^{\circ}$ to
study the origin of the temperature oscillations. This is because
in addition to the twisting and off-center sloshing motion, the
orientation of the LSC also meanders randomly. By tilting the cell
and thus locking the LSC orientation, one can remove the
stochastic meandering from the signal and separate the complicated
phenomena produced by the different types of motions. This enabled
us to study the phase relationships between temperature
oscillations at various locations. In this work, our focus is on
the relationship between the twisting motion near the top and
bottom plates and the off-center motion at the mid-height plane.
As the LSC meanders azimuthally as a whole across the height of
the cell \cite{sun2005prl}, the phase relationship between these
two types of motions can be studied even with the azimuthal
meandering present. Therefore, unless stated otherwise, all
measurements in the present work were made with the cell levelled
(to within $0\pm 0.06^{\circ}$).

As stated above, the temperature measurements in the present case
are made with thermistors embedded in the sidewall, whereas in
Ref. \cite{xi2008prl} the off-center motion was obtained from
thermistors placed in fluid. It is known that the sidewall acts as
a low-pass filter, thus the thermistors embedded in the sidewall
are not sensitive to the high-frequency signals. Spatially, they
actually sense the integrated signal over a finite area of the
sidewall, which leads to a lowered strength of the off-center
oscillation as compared from that using the in-fluid probes.
Nevertheless, as we will see below, the basic pictures obtained
from the two cases are essentially the same. As one of the
objectives of the present work is to understand the relationship
between the off-center oscillation and the twisting oscillations
near the plates, we use data measured simultaneously by the 24
in-wall probes from the three heights (there are only 8 in-fluid
probes and that measurements can only be made at one height at a
time). Thus, unless stated otherwise, all results presented in
this paper were obtained from the in-wall probes.

\subsection{The temperature-extremum-extraction method}

An often-used method based on the multi-probe technique for
extracting the dynamic information about the LSC motion is to fit
the temperature azimuthal profile using a sinusoidal function,
i.e. $T_k=T_a+A'cos(k\pi/4-\phi'), k=0, 1, ..., 7$, where $T_a$ is
the azimuthal average of the 8 temperature readings, $A'$ is a
measure of the strength of the LSC and $\phi'$ is the LSC's
orientation (see, e.g. \cite{cioni1997jfm, brown2005prl,
brown2006jfm, xi2007pre, xi2008pof, funfschilling2008jfm}). This
sinusoidal-fitting (SF) method has been very successful in the
study of the azimuthal motion of the LSC, including rotations,
cessations and reversals (see \cite{agl2008rmp} and references
therein). However, the SF method requires the separation between
the hottest and coldest azimuthal positions to be $\pi$, i.e., the
obtained LSC's central line is forced to always pass through the
cell's central vertical axis, which, as we shall show below, is
not always the case.

\begin{figure}
\center
\resizebox{1\columnwidth}{!}{%
  \includegraphics{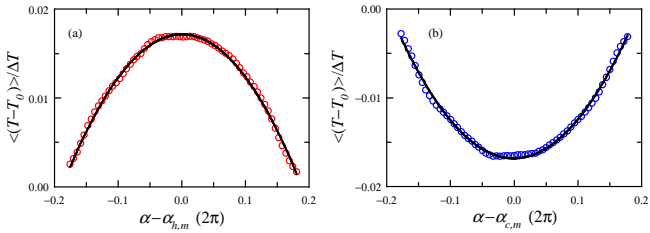}
}\caption{The averaged sidewall-temperature-profile
$[T(\alpha)-T_0]$ normalized by the temperature difference $\Delta
T$ around the fitted (a) hottest and (b) coldest azimuthal
positions obtained at $Ra=5.5\times10^9$. Solid lines are
quadratic functions.}\label{fig:fig2}
\end{figure}

Here, we introduce the temperature-extremum-extraction (TEE)
method that determines first the hottest and coldest azimuthal
positions of the bulk fluid and then the central line of the LSC
band. The TEE method has been described very briefly in Ref.
\cite{xi2008prl}. In this section, we give its detailed
description, validation and comparison with the SF method. To
illustrate this mehtod, figure \ref{fig:fig1}(c) shows an example
of the actual instantaneous temperature readings (circles) from
the 8 thermistors at mid-height plane. The hottest (coldest)
azimuthal position of the bulk fluid along the sidewall was
determined by making a quadratic fit around the highest (lowest)
temperature reading. In practice, this requires only the local
maximum (minimum) temperature and the two temperatures directly
adjacent to it. This is because a quadratic function has 3 degrees
of freedom and thus can be uniquely determined by only 3 data
points. We label these azimuthal positions $\alpha_1$, $\alpha_2$
and $\alpha_3$ with $\alpha_2$ being the position of the local
maximum (minimum) temperature. The position $\alpha_{h}$
($\alpha_{c}$) at any plane can then be found by solving
analytically the 3 quadratic equations between $T({\alpha_k})$ and
$\alpha_k$ ($k=1,2,3$), i.e.,
\begin{widetext}
\begin{equation}
\alpha_{h} (\alpha_{c}) = \frac{1}{2}
\frac{(\alpha_1^2-\alpha_2^2)[T(\alpha_2)-T(\alpha_3)]-(\alpha_2^2-\alpha_3^2)[T(\alpha_1)-T(\alpha_2)]}{(\alpha_1-\alpha_2)[T(\alpha_2)-T(\alpha_3)]-(\alpha_2-\alpha_3)[T(\alpha_1)-T(\alpha_2)]}.
\end{equation}
\end{widetext}
In this paper, the subscripts `$h$' and `$c$' are used to denote
those quantities associated with the hot ascending and cold
descending flows, respectively, and the subscripts `$t$', `$m$'
and `$b$' for the top, middle and bottom planes, respectively. To
test the validity of the quadratic fit, figure \ref{fig:fig2}
shows the normalized sidewall-temperature-profile. Each point is
an average of $[T(\alpha_k)-T_0]/\Delta T$ ($k=\alpha_1, \alpha_2,
\alpha_3$) in a bin with a small range around
$\alpha-\alpha_{h,m}$ (figure \ref{fig:fig2}(a)) or
$\alpha-\alpha_{c,m}$ (figure \ref{fig:fig2}(b)). The data around
the hottest (figure \ref{fig:fig2}(a)) and coldest (figure
\ref{fig:fig2}(b)) positions are both in good agreement with the
quadratic functions (solid lines), indicating that the quadratic
function is indeed a good representation for the temperature
distribution around the hot ascending and cold descending flows.
With this method, the three rows of thermistors can thus provide
simultaneously the azimuthal positions of the hot ascending and
cold descending flows at the three heights, which are denoted as
$\alpha_{h,t}$, $\alpha_{h,m}$, and $\alpha_{h,b}$, and
$\alpha_{c,t}$, $\alpha_{c,m}$, and $\alpha_{c,b}$. With the
obtained hottest and coldest positions, the line connecting the
two positions is the central line of the LSC band. The orientation
of this line is the orientation of the LSC, the distance between
this line and the cell's central vertical axis is defined as the
off-center distance (see figure \ref{fig:fig1}(a)) and
$A=(T(\alpha_h)-T(\alpha_c))/2$ is used to characterize the
strength of the LSC, where the factor 1/2 is used in order to
compare the obtained $A$ with that obtained from the SF method. We
denote $\phi_t$, $\phi_m$, and $\phi_b$, $A_t$, $A_m$, and $A_b$,
and $d_t$, $d_m$, and $d_b$ as the orientations, the strengths and
the off-center distances of the LSC at the top, middle, and bottom
planes, respectively. Therefore, based on the TEE method, the
motion of the LSC can be decomposed into two different modes: The
azimuthal mode and the off-center mode.

\begin{figure}
\center
\resizebox{1\columnwidth}{!}{%
  \includegraphics{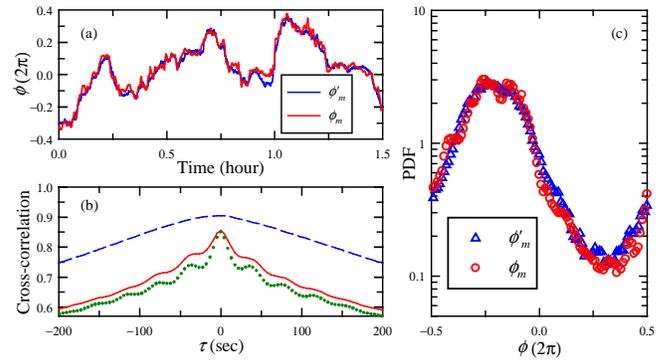}
}\caption{(a) Time traces of the orientation $\phi_m$ of the LSC
central line obtained using the TEE method and the orientation
$\phi'_m$ obtained from the SF method. (b) The cross-correlation
functions between the orientations obtained using the two
different methods for the top (solid line), middle (dashed line)
and bottom (dotted line) planes. (c) PDF of the two orientations
$\phi_m$ and $\phi'_m$. All data were measured at
$Ra=5.5\times10^9$.}\label{fig:fig3}
\end{figure}

We now compare the orientations and the strengths of the LSC
obtained using the TEE method and those obtained using the SF
method. Figure \ref{fig:fig3}(a) shows the time traces of $\phi_m$
and $\phi'_m$, which are obtained using the same data set measured
by the embedded thermistors. One sees that the time traces are
very similar to each other. This similarity can be characterized
by the coefficient of the cross-correlation between the two
quantities. The cross-correlation function between two variables
$a$ and $b$ is defined as
\begin{equation}
C_{a,b}(\tau)=\langle (a(t+\tau)-\langle a\rangle)(b(t)-\langle
b\rangle) \rangle/a_{rms}b_{rms},
\end{equation}
where $a_{rms}$ and $b_{rms}$ are standard deviations of $a$ and
$b$, respectively, and $\langle\cdots\rangle$ represents the
temporal average. When $a=b$, $C_{a,a}$, denoted as $C_a$, is the
auto-correlation function of the variable $a$. Figure
\ref{fig:fig3}(b) shows the cross-correlation functions between
$\phi$ and $\phi'$ at the three heights. It shows that all these
three functions have a positive and strong correlation with the
coefficient being all larger than 0.85 at time lag $\tau=0$. (The
subpeaks of $C_{\phi_t,\phi'_t}$ and $C_{\phi_b,\phi'_b}$
correspond to the twisting oscillations of the LSC near the top
and bottom plates.) It is further found that this strong positive
correlation exists for all values of $Ra$ investigated. Figure
\ref{fig:fig3}(c) shows the probability density functions (PDF) of
$\phi_m$ and $\phi'_m$. One sees that the two PDFs essentially
collapse on top of each other except that the distribution of
$\phi'_m$ seems to be a little smoother than that of $\phi_m$.

\begin{figure}
\center
\resizebox{1\columnwidth}{!}{%
  \includegraphics{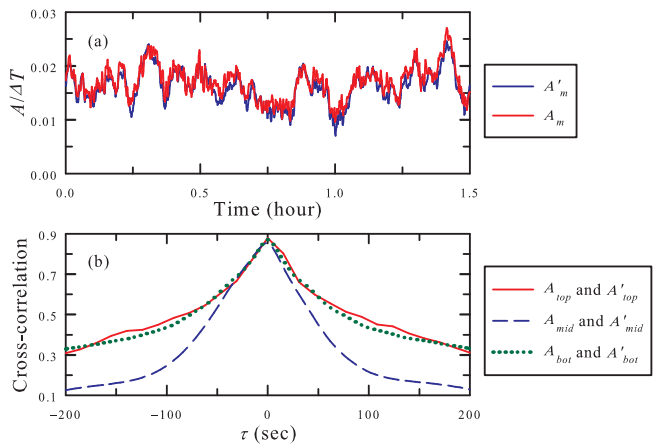}
}\caption{(a) Time traces of the normalized strength $A_m/\Delta
T$ of the LSC obtained using the TEE method and the normalized
strength $A'_m/\Delta T$ obtained from the SF method. (b) The
cross-correlation functions between the strengths obtained using
the two different methods for the top (solid line), middle (dashed
line) and bottom (dotted line) planes. The data were measured at
$Ra=5.5\times10^9$.}\label{fig:fig3a}
\end{figure}

Figure \ref{fig:fig3a}(a) shows the time traces of $A_m$ and
$A'_m$. It is seen that the two time traces are again very similar
to each other except that $A_m$ is on average $\sim5\%$ larger
than $A'_m$. Figure \ref{fig:fig3a}(b) shows the cross-correlation
functions between $A$ and $A'$ at the three heights. It similarly
shows that all three functions have a strong positive correlation
with their peaks all located at time lag $\tau=0$ and the
cross-correlation coefficient $C(\tau=0)$ all being nearly 0.9.

\begin{figure}
\center
\resizebox{1\columnwidth}{!}{%
  \includegraphics{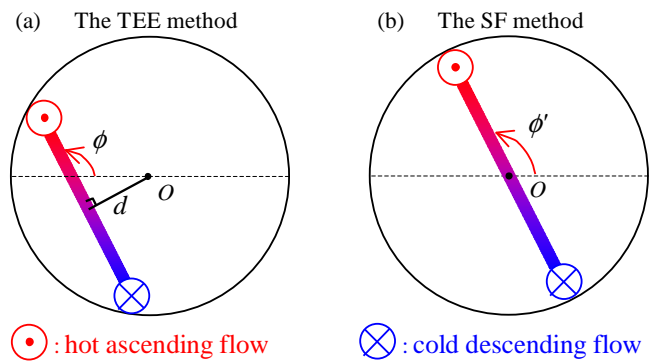}
}\caption{(a) Schematic diagram showing the definitions of the
orientation $\phi$ and the off-center distance $d$ of the LSC
obtained using the TEE method. (b) Schematic diagram showing the
definition of the orientation $\phi'$ of the LSC obtained using
the SF method. In this method, the off-center distance is forced
to be always zero.}\label{fig:fig3b}
\end{figure}

The above results imply that, as far as the orientation and the
strength of the LSC are concerned, both the SF and TEE methods
give the same results, but the TEE method in addition determines
the LSC's off-center motion, which is missed by the SF method.
This is illustrated in figures \ref{fig:fig3b}(a) and (b). It is
also clear that the LSC's orientation is determined mainly by the
hottest and coldest azimuthal positions of the bulk fluid along
the sidewall. This is not surprising, since one can see from
figure \ref{fig:fig1}(c) that the sinusoidal fit to the
temperature profile shifts the hottest azimuthal position
$\alpha_{h,m}$ right or anticlockwise and shifts the coldest
azimuthal position $\alpha_{c,m}$ left or clockwise and the
changes of the obtained orientation of the LSC due to the two
shifts roughly cancel each other.

\begin{figure}
\center
\resizebox{1\columnwidth}{!}{%
  \includegraphics{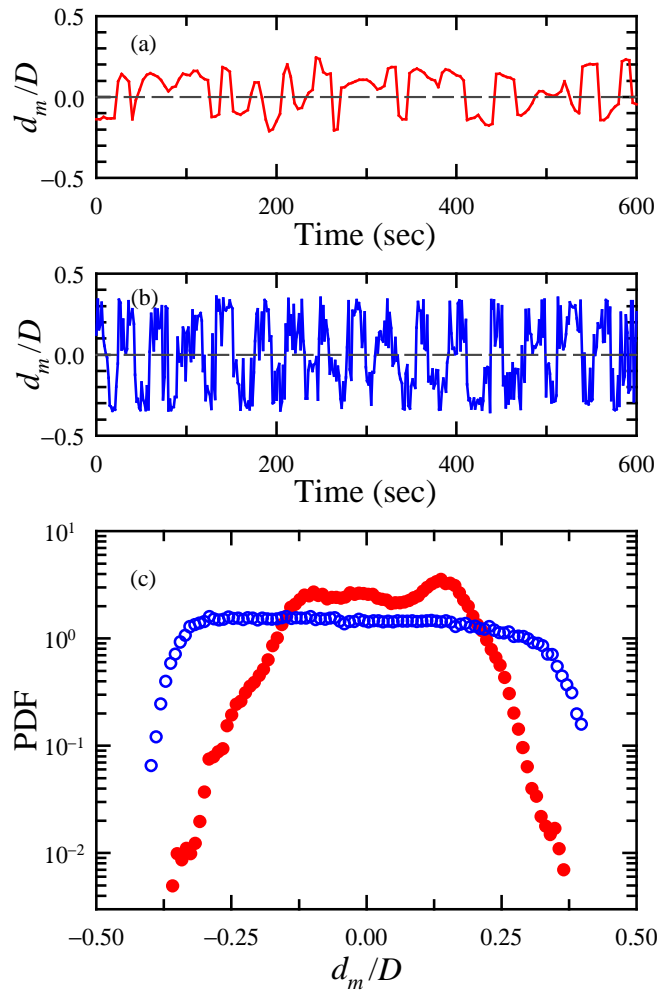}
}\caption{(a, b) Time traces of the normalized off-center distance
$d_m/D$ from thermistors (a) embedded into the sidewall and (b)
placed in fluid. (c) PDF of $d_m/D$ from the in-wall (solid
circles) and in-fluid (open circles) thermistors. The data of the
in-fluid thermistors are taken at $Ra=5.0\times10^9$ with a
sampling rate of 0.76 Hz and that of the in-wall probes at
$Ra=5.5\times10^9$ with a sampling rate of 0.29
Hz.}\label{fig:fig4}
\end{figure}

\section{Results and discussion}

\subsection{The off-center oscillation of the LSC at mid-height plane}

The off-center distance $d_m$ of the LSC's central line at the
mid-height plane is used to study the off-center motion of the LSC
at that plane. Figure \ref{fig:fig4}(a) shows a typical time trace
of the measured $d_m$ normalized by the cell's diameter $D$. It is
seen from the figure that $d_m/D$ fluctuates around 0. We further
found that the temporal averages of $d_m/D$ for all six values of
$Ra$ are nearly 0. In Ref. \cite{xi2008prl} a similar trace
measured with the in-fluid thermistors in a tilted cell has been
shown. For comparison, we plot in figure \ref{fig:fig4}(b) a time
trace of $d_m/D$ measured under the same conditions as in Ref.
\cite{xi2008prl} but with the cell leveled ($0\pm0.06^{\circ}$).
It shows that $d_m/D$ measured from the in-fluid probes has a
stronger strength and its oscillation looks more periodic when
compared with that obtained from the in-wall thermistors.
Nevertheless, the physical pictures revealed by the two time
traces are the same, i.e., the LSC's central line oscillates
horizontally around the cell's central vertical axis. Figure
\ref{fig:fig4}(c) shows the PDF of $d_m/D$ obtained from the
in-wall thermistors (solid circles). The PDF is found to be
flatter than a Gaussian function, i.e. the flatness is $2.3$. The
PDFs obtained at other $Ra$ share the same features as that shown
in figure \ref{fig:fig4}(c), except for the two lowest $Ra$ which
is probably due to the limited temperature contrast along the
sidewall. For comparison, we also plot the PDF of $d_m/D$ obtained
from the in-fluid thermistors (open circles). It is seen that the
PDF from the in-fluid probes is even flatter (flatness 1.9) than
that from the in-wall thermistors, reflecting a larger range of
oscillation.

\begin{figure}
\center
\resizebox{1\columnwidth}{!}{%
  \includegraphics{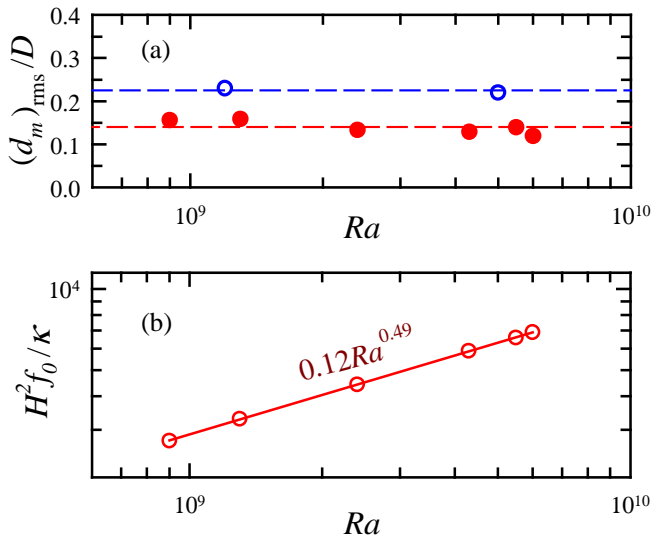}
}\caption{(a) The $Ra$-dependence of the standard deviation of
$d_m/D$ from the in-wall (solid circles) and in-fluid (open
circles) thermistors. Two dashed lines show the mean value of
$d_m/D$ averaged over all $Ra$ investigated. (b) Normalized
oscillation frequency $H^2f_0/\kappa$ obtained from the frequency
power spectra of $d_m$ as a function of $Ra$. The solid line is
the power-law fit.}\label{fig:fig4a}
\end{figure}

As $d_m/D$ oscillates around its mean value of 0, its standard
deviation $(d_m)_{rms}/D$ can be used as a measure of the
amplitude of this oscillation. Figure \ref{fig:fig4a}(a) shows the
$Ra$-dependence of $(d_m)_{rms}/D$ obtained from the in-wall
(solid circles) and in-fluid (open circles) thermistors. One sees
that $(d_m)_{rms}/D$ depends weakly on $Ra$ for the present range
of $Ra$ with $(d_m)_{rms}/D$ from the in-fluid probes being
$\sim60\%$ larger than that obtained from the in-wall thermistors.
In Ref. \cite{xi2008prl} it has been shown that the off-center or
sloshing motion of the LSC exhibits a well-defined time-periodic
oscillation. The prominent peak near $f_0$ in the frequency power
spectra of $d_m$ (see the second curve from bottom in figure
\ref{fig:fig5}(a)) corresponds to this periodic oscillation. In
figure \ref{fig:fig4a}(b) we plot the $Ra$-dependence of the
normalized frequency corresponding to this prominent peak obtained
from the spectra of $d_m$. The solid line in the figure represents
the power-law fit to the data:
$H^2f_0/\kappa=0.12Ra^{0.49\pm0.02}$, with the scaling exponent
being in good agreement with the previous temperature measurements
in water \cite{qiu2001prl, qiu2002pre, sunxia2005pre,
brown2007jstat}, mercury \cite{takeshita1996prl} and
low-temperature helium gas \cite{heslot1987pra, castaing1989jfm,
sano1989pra, niemela2001jfm}. This result further indicates that
the previously-observed temperature oscillations near the
mid-height of the cell indeed originates from the off-center
oscillation of the LSC. Taken together, the physical picture
behind figure \ref{fig:fig4} is that the LSC's central line
oscillates time-periodically around the cell's central vertical
axis with an amplitude nearly independent of the turbulent
intensity.

\subsection{The relationship between the off-center mode and the torsional mode of the LSC}

\begin{figure*}
\begin{minipage}[c]{.75\textwidth}
\centering
\includegraphics[trim=2cm 0cm 0cm 0cm, scale=0.8]{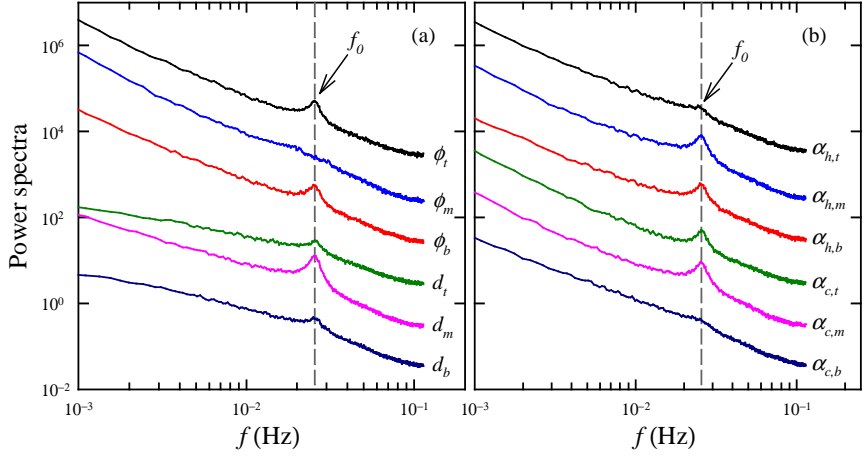}
\end{minipage}
\centering
\begin{minipage}[c]{.2\textwidth}
\centering \caption{From top to bottom: Power spectra of (a)
$\phi_t$, $\phi_m$, $\phi_b$, $d_t$, $d_m$, and $d_b$, and (b)
$\alpha_{h,t}$, $\alpha_{h,m}$, $\alpha_{h,b}$, $\alpha_{c,t}$,
$\alpha_{c,m}$, and $\alpha_{c,b}$. All data have been shifted
vertically for clarity. \label{fig:fig5}}
\end{minipage}
\end{figure*}

Figure \ref{fig:fig5}(a) shows, from top to bottom, the frequency
power spectra of $\phi_t$, $\phi_m$, $\phi_b$, $d_t$, $d_m$, and
$d_b$. For the orientations of the LSC obtained at the three
heights, the prominent peak near $f_0$ representing periodic
oscillation can be seen clearly for $\phi_t$ and $\phi_b$, but is
absent for the mid-height plane ($\phi_m$). These are consistent
with those observed by Ref. \cite{funfschilling2008jfm} and
correspond to the twisting mode of the LSC
\cite{funfschilling2004prl}. We further found that although
$\phi_m$ will also exhibit oscillation when the cell is tilted by
several degrees, the strength of this oscillation obtained at the
mid-height plane is much weaker than those of $\phi_t$ and
$\phi_b$, i.e., the peak height of the spectra of $\phi_m$ at
$f_0$ is much smaller than those of $\phi_t$ and $\phi_b$.
However, for the off-center distance, the situation is opposite.
In figure \ref{fig:fig5}(a) one sees that the strength of the
oscillation peak for $d_m$ at the mid-height plane is much
stronger than those of $d_t$ and $d_b$.

\begin{figure*}
\begin{minipage}[c]{.75\textwidth}
\centering
\includegraphics[trim=2cm 0cm 0cm 0cm, scale=0.85]{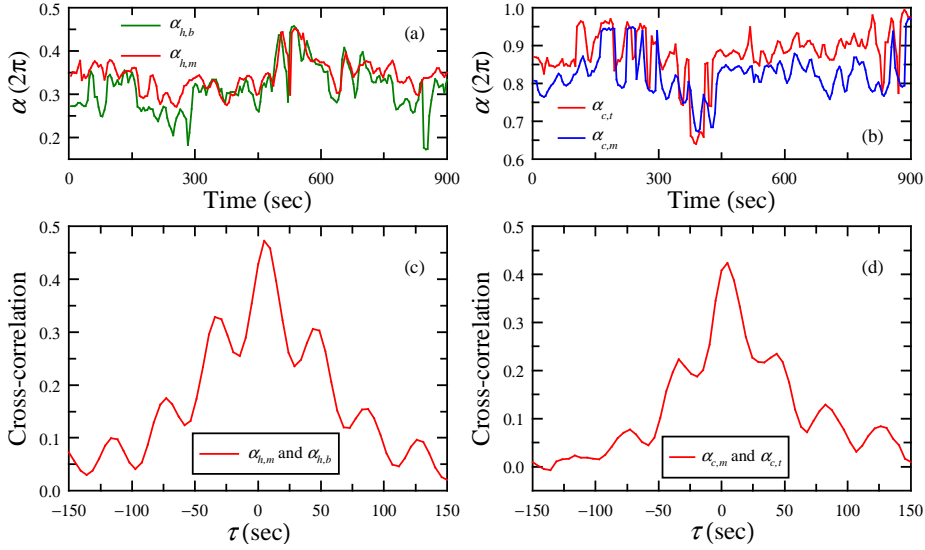}
\end{minipage}
\centering
\begin{minipage}[c]{.2\textwidth}
\centering \caption{Time traces of (a) $\alpha_{h,b}$ and
$\alpha_{h,m}$ and of (b) $\alpha_{c,t}$ and $\alpha_{c,m}$. The
cross-correlation functions between (c) $\alpha_{h,m}$ and
$\alpha_{h,b}$ and between (d) $\alpha_{c,m}$ and $\alpha_{c,t}$.
All data were obtained at $Ra=5.5\times10^9$. \label{fig:fig6}}
\end{minipage}
\end{figure*}

To understand this difference and to find out the relationship
between the off-center mode at the mid-height plane and the
torsional mode of the LSC near the top and bottom plates, we
examine the frequency power spectra of the azimuthal positions of
the hot ascending and cold descending flows at the three heights,
as both the orientation $\phi$ and the off-center distance $d$ of
the LSC are obtained from these positions. Figure
\ref{fig:fig5}(b) shows that the spectra of $\alpha_{c,t}$ and
$\alpha_{h,b}$ exhibit an oscillation peak near $f_0$. Whereas the
spectra of $\alpha_{h,t}$ and $\alpha_{c,b}$ both give very faint
oscillations, which is due to a tilted ellipselike circulation
path of the LSC when viewed from the side (Qiu $\&$ Tong 2001b;
Sun $\&$ Xia 2005; Sun \emph{et al.} 2005b). The thermistors
embedded into the sidewall can not feel accurately the hot
ascending flow of the LSC at the top-plane and the cold descending
flow of the LSC at the bottom-plane, since they are far away from
the sidewall at the respective heights (figure \ref{fig:fig1}(b)).
Therefore, at the top and bottom planes, the in-wall probes may
not be able to measure the orientation and the off-center distance
of the LSC as accurately as those at the mid-height plane. This
suggests that the relatively weak oscillation peaks of $d_t$ and
$d_b$ shown in figure \ref{fig:fig5}(a) are mainly due to the
oscillations of $\alpha_{c,t}$ and $\alpha_{h,b}$. The surprising
thing shown in figure \ref{fig:fig5}(b) is that the spectra of the
azimuthal positions $\alpha_{h,m}$ and $\alpha_{c,m}$ of the hot
ascending and cold descending flows at the \emph{mid-height plane}
both exhibit a well-defined time-periodic oscillation with the
oscillation peak located at $f_0$ and with the same strength as
those of $\alpha_{h,b}$ and $\alpha_{c,t}$. This implies that the
azimuthal positions of the hot ascending and cold descending
fluids both oscillate periodically along the sidewall irrespective
of whether they are near the plates or at the mid-height of the
cell.

\begin{figure}
\center
\resizebox{1\columnwidth}{!}{%
  \includegraphics{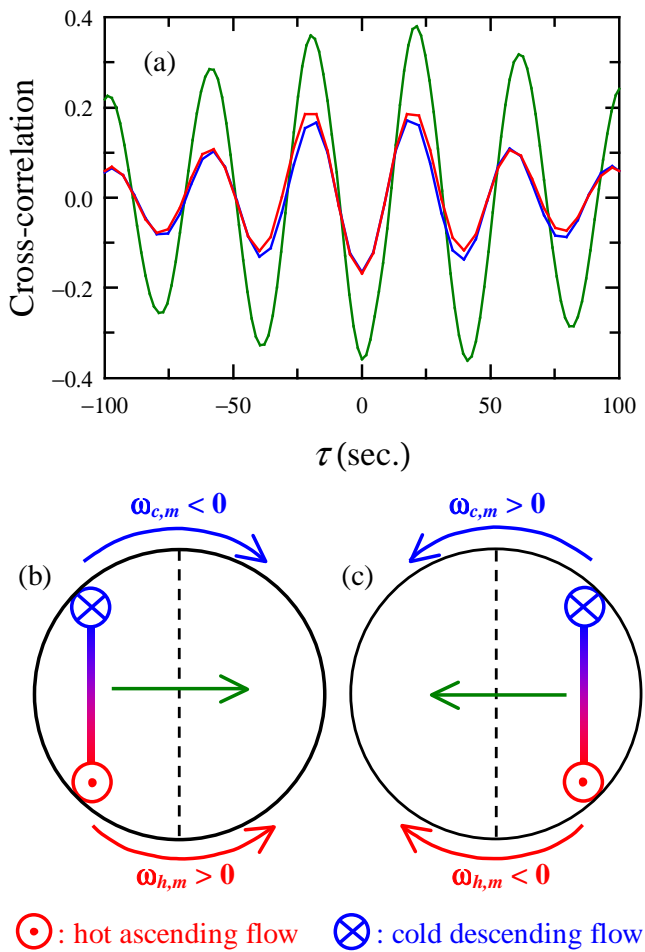}
}\caption{(a) The cross-correlation functions
$C_{\omega_{c,t},\omega_{h,b}}$ (blue line) and
$C_{\omega_{c,m},\omega_{h,m}}$ from the in-wall (red line) and
in-fluid thermistors (dark green line). The data from in-wall
thermistors are taken at $Ra=5.5\times10^9$ and that from in-fluid
probes at $Ra=5.0\times10^9$. (b) Schematic diagram of the
azimuthal positions of the hot ascending and cold descending flows
when the off-center oscillation is at an extremum. The curved
arrows show the directions of the angular velocities and the
horizontal arrow shows the moving direction of the LSC's central
line. (c) Same as (b) but half an oscillation period later.
}\label{fig:fig7}
\end{figure}

Figure \ref{fig:fig6}(a) shows typical time traces of
$\alpha_{h,b}$ and $\alpha_{h,m}$ measured at the same time. It is
seen that the two time traces are rather similar to each other
except for a several-second delay between them. The
cross-correlation function $C_{\alpha_{h,m},\alpha_{h,b}}$ shown
in figure \ref{fig:fig6}(c) quantifies this similarity. It shows
that $\alpha_{h,m}$ and $\alpha_{h,b}$ have a strong positive
correlation with the main peak located at $\tau\simeq6\pm1$ s.
This positive time-delay indicates that the azimuthal motion of
$\alpha_{h,m}$ lags that of $\alpha_{h,b}$, which is easy to
understand since the hot ascending flow rises up from bottom.
Similar situation can be seen for the relation between
$\alpha_{c,t}$ and $\alpha_{c,m}$. Figure \ref{fig:fig6}(b) shows
the time traces of $\alpha_{c,t}$ and $\alpha_{c,m}$, which again
are quite similar to each other. Figure \ref{fig:fig6}(d) plots
$C_{\alpha_{c,m},\alpha_{c,t}}$, and it is found that
$\alpha_{c,m}$ correlates strongly with $\alpha_{c,t}$ with a
positive time delay $\tau\simeq5\pm1$ s indicating that the
azimuthal motion of $\alpha_{c,m}$ lags that of $\alpha_{c,t}$.
Here we note that the time delay of the main peak of
$C_{\alpha_{c,m},\alpha_{c,t}}$ is essentially the same as that of
$C_{\alpha_{h,m},\alpha_{h,b}}$ within experimental uncertainty,
this is due to the fact that the distance between the middle and
the top planes is the same as the distance between the middle and
bottom ones. Note also that the subpeaks of
$C_{\alpha_{h,m},\alpha_{h,b}}$ and
$C_{\alpha_{c,m},\alpha_{c,t}}$ correspond to the periodic
oscillations shown in figure \ref{fig:fig5}(b). These results
further imply that the hot ascending and cold descending flows go
up and fall down coherently, thus propagate their azimuthal
positional oscillations along the sidewall.

The torsional mode of the LSC can be revealed by the horizontal
motions of the hot ascending and cold descending flows near the
top and bottom plates. Here, we define
\begin{equation}
\omega_{i,j} = d\alpha_{i,j}/dt,
\end{equation}
as the azimuthal angular velocity of $\alpha_{i,j}$, where $i=h$
or $c$ and $j=t$, $m$, or $b$. Figure \ref{fig:fig7}(a) shows the
cross-correlation function $C_{\omega_{c,t},\omega_{h,b}}$ between
$\omega_{c,t}$ and $\omega_{h,b}$ (blue line). It is seen that
$C_{\omega_{c,t},\omega_{h,b}}$ oscillates and $\omega_{c,t}$
anticorrelates with $\omega_{h,b}$ with a strong negative peak
located at $\tau=0$. The negative peak at $\tau=0$ indicates that
$\alpha_{c,t}$ rotates clockwise when $\alpha_{h,b}$ rotates
anticlockwise and vice versa, implying the twisting motion of the
LSC. Recall that the azimuthal motions of $\alpha_{h,m}$ and
$\alpha_{c,m}$ follow coherently those of $\alpha_{h,b}$ and
$\alpha_{c,t}$, respectively, with the same time delay. The
azimuthal angular velocities of $\alpha_{h,m}$ and $\alpha_{c,m}$
would hence be expected to exhibit the same behavior as those of
$\alpha_{h,b}$ and $\alpha_{c,t}$, and this is indeed observed
from figure \ref{fig:fig7}(a), which shows that
$C_{\omega_{c,m},\omega_{h,m}}$ (red line) is the same as
$C_{\omega_{c,t},\omega_{h,b}}$ (blue), i.e. periodic oscillations
and a strong negative peak at $\tau=0$. Here, we also plot
$C_{\omega_{c,m},\omega_{h,m}}$ obtained from the in-fluid
thermistors (dark green line). It is seen that
$C_{\omega_{c,m},\omega_{h,m}}$ obtained from the in-fluid
thermistors exhibits the oscillation with much stronger
periodicity than that obtained from the in-wall thermistors (red
line). However, the basic picture is the same for both cases.
These results indicate that the azimuthal positional oscillations
of the hot ascending and cold descending flows at the mid-height
plane are out of phase with each other by $\pi$, leading to the
off-center mode of the LSC at the mid-height plane. The mechanism
of how this off-center mode is generated is illustrated in figures
\ref{fig:fig7}(b) and (c). Figure \ref{fig:fig7}(b) shows that
$\alpha_{c,m}$ rotates clockwise when $\alpha_{h,m}$ rotates
anticlockwise, and the line connecting them thus moves to the
right. Half an oscillation period later (figure
\ref{fig:fig7}(c)), the motion of $\alpha_{c,m}$ changes to
anticlockwise while that of $\alpha_{h,m}$ to clockwise. This
makes the LSC central line move to the left. The periodic
occurrence of this process thus generates the off-center
oscillation of the LSC and produces the prominent peak near $f_0$
on the spectra of $d_m$. On the other hand, as the orientation
$\phi_m$ equals to $(\alpha_{h,m}+\alpha_{c,m})/2$ plus a
constant, the anticorrelation between $\omega_{h,m}$ and
$\omega_{c,m}$ would cancel out the periodic oscillations, and
hence it is not surprising to see in figure \ref{fig:fig5}(a) that
an oscillation peak is absent for the spectra of $\phi_m$.

\subsection{Direct velocity measurement of the off-center motion}

\begin{figure*}
\begin{minipage}[c]{.7\textwidth}
\centering
\includegraphics[trim=2cm 0cm 0cm 0cm, scale=0.65]{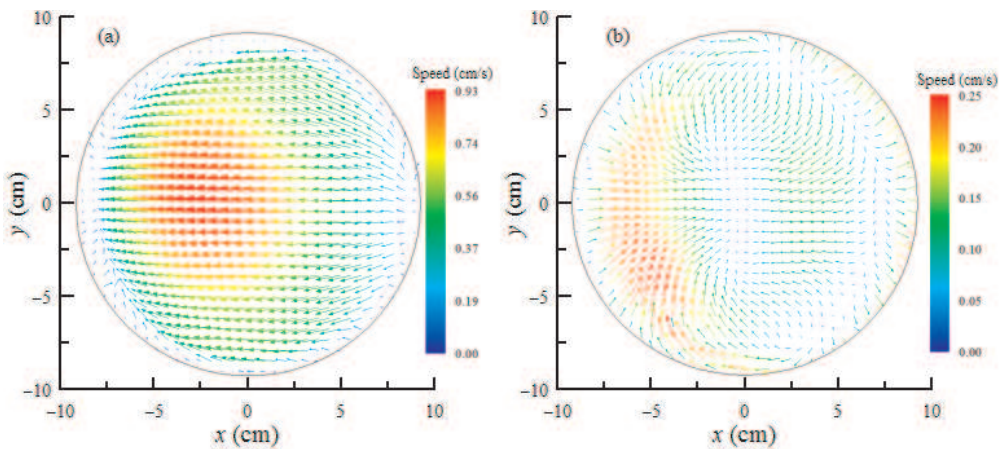}
\end{minipage}
\centering
\begin{minipage}[c]{.25\textwidth}
\centering \caption{Time-averaged velocity vector map measured (a)
near the top plate and (b) at the mid-height plane. For clarity, a
coarse-grained vector map of size $32\times32$ is shown. The
magnitude of the velocity $\sqrt{u^2+v^2}$ is coded in both the
color and the length of the arrows in units of cm/s. Both
time-averages are taken over a period of 151 min corresponding to
20000 velocity maps. \label{fig:fig14}}
\end{minipage}
\end{figure*}

We present in this subsection direct evidence of the off-center
oscillation of the LSC from particle image velocimetry (PIV)
measurement of the horizontal velocity field at the mid-height
plane in a sapphire cell. Both the sapphire cell and the
horizontal velocity measurement using the PIV technique have been
described in detail previously \cite{xi2006pre, zhou2007prl} and
hence we outline only their main features here. The sapphire cell
consists of two sapphire discs with thickness 5 mm as the top and
bottom plates and a Plexiglas tube with thickness 8 mm as the
sidewall. The cell's inner diameter and height are both 18.5 cm,
so the aspect ratio of the cell is also unity. In the present
work, 50-$\mu$m-diameter polyamid spheres (density 1.03
gcm$^{-3}$) were used as the seeding particles and a horizontal
laser light-sheet with thickness $\sim2$ mm was used to illuminate
the particles in the mid-height plane of the cell. The measuring
area is a square of $18.5\times18.5$ cm$^2$, which covers
completely the horizontal cross-section of the cell. The spatial
resolution is 0.29 cm, corresponding to $63\times63$ velocity
vectors in each 2D velocity map. Denote the laser-illuminated
plane as the $(x, y)$-plane and the center of the mid-height
horizontal cross-section of the cell as the origin $O$ of the
coordinates (see figure \ref{fig:fig12}(a)), then two horizontal
velocity components $u(x, y)$ and $v(x, y)$ are measured. In these
PIV measurements, the cell was tilted by $\sim1^{\circ}$ at
position 1, and thus the orientation of the LSC was locked along
the $x$-direction (see figure \ref{fig:fig1}(a)). For the PIV
experiment, the measurements were made at $Ra=3.0\times10^9$ and
$6.0\times10^9$ and at $Pr=4.3$. A total of 20000 vector maps were
acquired for each $Ra$ at a sampling rate of $\sim2.2$ Hz. As the
two measurements give the same results, only results for
$Ra=6.0\times10^9$ will be presented.

\begin{figure}
\center
\resizebox{1\columnwidth}{!}{%
  \includegraphics{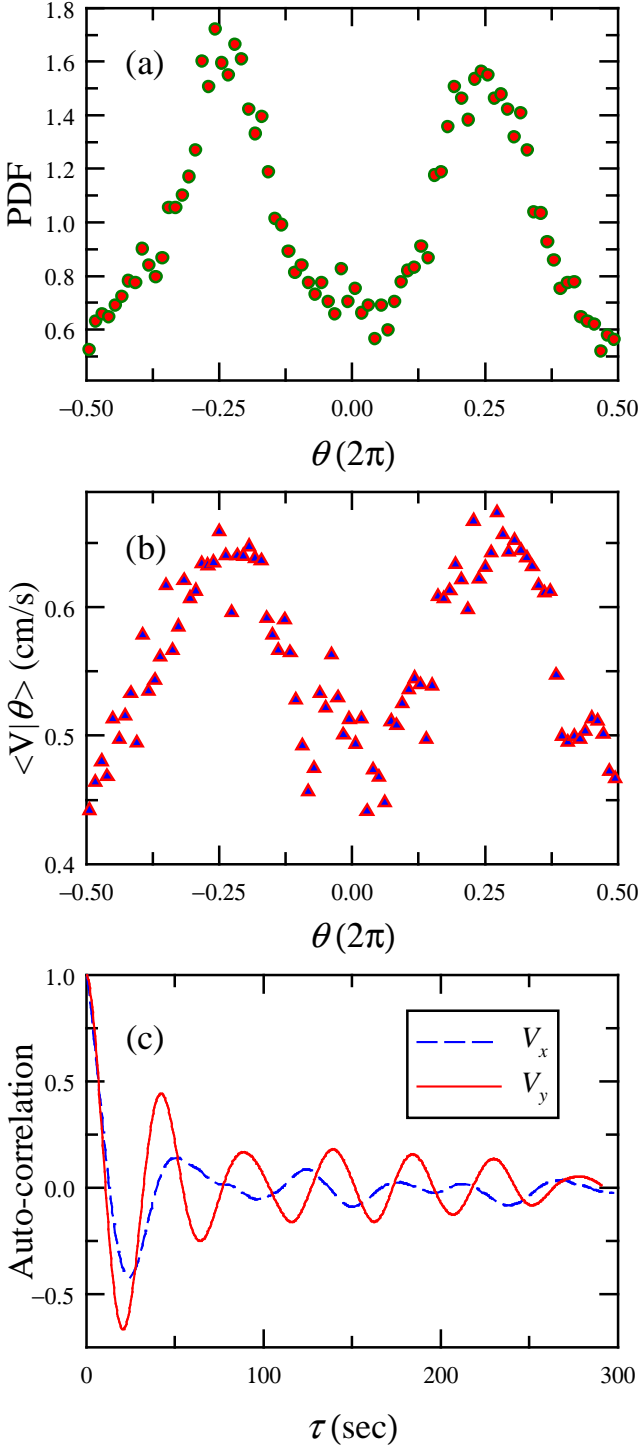}
}\caption{(a) PDF of the orientation $\theta$ of the spatially
averaged vector $\vec{V}$. (b) The conditional average $\langle
V|\theta \rangle$ on the velocity orientation $\theta$. (c) The
auto-correlation functions of $V_x$ and $V_y$.}\label{fig:fig8}
\end{figure}

To study the global motion of the central bulk fluid, we use the
velocity vector $\vec{V}(t)=V_x(t)\hat{x}+V_y(t)\hat{y}$ to
characterize the overall flow behavior at the mid-height plane,
where $V_x(t)$ and $V_y(t)$ are the spatial averages of the local
velocity components $u(t)$ and $v(t)$, respectively, within a
circular area centered at the center of the measurement area. The
diameter of the circular area is 5 cm and there are 221 vectors
contained in this circular region for averaging. A circle with a
diameter of 10 cm has also been used and the obtained results are
essentially the same \cite{xi2006pre}. The orientation
$\theta\equiv\arctan(V_y/V_x)$ and the magnitude
$V\equiv|\vec{V}|=\sqrt{V_x^2+V_y^2}$ of the averaged vector are
then measures of the orientation and the strength of the motion of
the central bulk fluid, respectively.

Note that the horizontal velocity field investigated in Ref.
\cite{xi2006pre} was obtained near the top plate. Figure
\ref{fig:fig14}(a) shows the time-averaged velocity vector map
measured at 2 cm from the top plate, which shows that the LSC is
dominated mainly by the horizontal velocities. Thus, the spatial
averaged velocity vector $\vec{V}(t)$ measured at the horizontal
planes near the top plate can be used as a measure of the
orientation vector $\vec{V}_{LSC}$ of the LSC \cite{xi2006pre} and
its orientation $\theta$ can be used to characterize the
orientation of the LSC at the corresponding height. However, the
situation is different when the measurement is made near the
cell's mid-height plane. Figure \ref{fig:fig14}(b) shows the
time-averaged velocity vector map obtained at the mid-height
horizontal plane of the cell. Because at the cell's mid-height
plane the LSC is concentrated near the sidewall and dominated
mainly by vertical velocities (see the circulation path of the LCS
shown in figure \ref{fig:fig1}(b)), the velocity vector
$\vec{V}(t)$ spatially averaged from the velocities in the central
bulk region cannot be used to represent the orientation vector of
the LSC and its orientation $\theta$ is no longer a direct measure
of the LSC's orientation. Therefore, $\theta$ and $\phi_m$ (the
orientation of the LSC obtained from the azimuthal temperature
profile) would exhibit different behaviors. As we shall see below,
this is indeed the case.

Figure \ref{fig:fig8}(a) shows the PDF of the measured velocity
orientation $\theta$. It is seen that the PDF of $\theta$, which
is significantly different from that of $\phi_m$ in figure
\ref{fig:fig3}(c), exhibits a bimodal distribution with two peaks
located at the orientations ($\theta=\pm0.25$) that are
perpendicular to the preferred orientation of the LSC
($\theta=0$). The probabilities of $\theta=\pm0.25$ are nearly 3
times larger than those of $\theta=0$ and $\theta=\pm0.5$,
suggesting that the central bulk fluid is much more likely to move
in the direction perpendicular to the LSC's vertical circulation
plane rather than in the LSC's preferred direction itself. To
study the flow strength of the central bulk fluid in different
orientations, we calculate the conditional average $\langle
V|\theta\rangle$ on the velocity orientation $\theta$, as shown in
figure \ref{fig:fig8}(b). One sees that the averaged velocity
magnitudes in the directions ($\theta=\pm0.25$) perpendicular to
the LSC plane are much stronger than the magnitudes in all other
directions, especially than that in the preferred direction of the
LSC. Figure \ref{fig:fig8}(c) shows the auto-correlation functions
$C_{V_x}$ (dashed line) and $C_{V_y}$ (solid line). Both show
oscillations, but the oscillation strength of $V_y$ (in the
direction perpendicular to the LSC plane) is much stronger than
that of $V_x$ (along the LSC's preferred direction) and $V_y$
oscillates more coherently than $V_x$. In fact, a previous
single-point velocity measurement has shown that at the cell
center the strongest velocity oscillation is along the direction
perpendicular to the LSC plane and that the strength of this
oscillation decays away from the cell center towards the plates
\cite{qiu2004pof, sun2005pre}. We further note that the
oscillation frequency of $V_y$ is the same as that of the
off-center motion of the LSC but $20\%$ larger than that of $V_x$.
The existence of two different oscillation frequencies in
turbulent convection with fixed control parameters ($Ra$ and $Pr$)
have already been observed previously \cite{xi2006pre,
brown2007jstat}, but the reason(s) for the presence of two clocks
in turbulent RB system remain unknown \cite{xia2007jstat}. Here we
offer a possible reason for the observed different oscillation
frequencies of $V_x$ and $V_y$. Even though the convection cell
has been tilted by $\sim1^{\circ}$ in the PIV measurement, the
orientation of the LSC can still meander around the locked
direction (the $x$-axis in the present case) over an azimuthal
angular range of a few tens degrees \cite{ahlers2006jfm}. The
projection of the off-center oscillation onto the $x$-direction
would thus generate the oscillation in $V_x$, while the stochastic
fluctuations of the projection angle would both reduce the
coherence and change the frequency of the oscillation of $V_x$.
Note that the different oscillation frequencies for $V_x$ and
$V_y$ have also been observed, but not explicitly recognized, by
Refs. \cite{qiu2004pof} and \cite{sun2005pre}. Notwithstanding the
above, the overall picture emerging from the PIV measurements is
that the translational motion of the central bulk fluid in the
direction perpendicular to the LSC's vertical circulation plane is
more probable, much stronger and more coherent than the motions in
all other directions, which provides a direct evidence for the
off-center motion of the LSC at the mid-height plane.

\begin{figure*}
\begin{minipage}[c]{.65\textwidth}
\centering
\includegraphics[trim=2cm 0cm 0cm 0cm, scale=0.75]{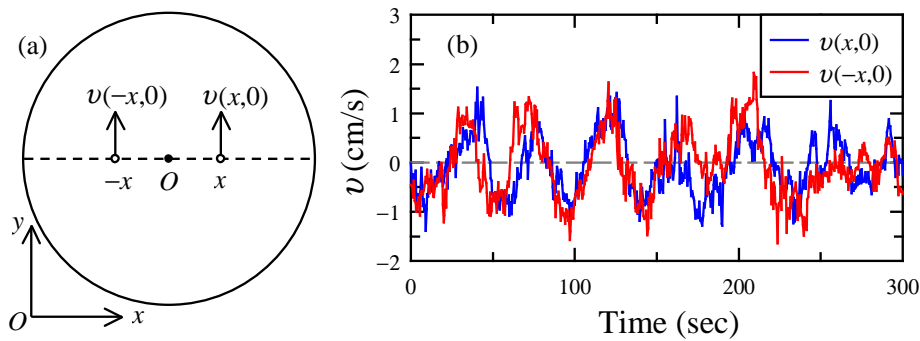}
\end{minipage}
\centering
\begin{minipage}[c]{.32\textwidth}
\centering \caption{(a) Coordinate system and schematic diagram of
the mid-height horizontal cross-section of the cell. The origin
$O$ of the coordinates is chosen as the center of the
cross-section and thus the $z$-axis (not shown in the figure) is
along the cell's central vertical axis. The two vertical arrows
show the velocity components $v$ located at positions $(x,0)$ and
$(-x,0)$ and the dashed line shows the $x$-axis. (b) Time traces
of $v(x,0)$ and $v(-x,0)$ for $x/D=0.11$. \label{fig:fig12}}
\end{minipage}
\end{figure*}

In a previous work, Sun \emph{et al.} \cite{sun2005pre} measured
the 2D velocity field in the $(y, z)$-plane, which is
perpendicular to the LSC's vertical circulation plane and studied
the phase relationships among the horizontal velocity components
$v(y)$ along the $y$-axis and among $v(z)$ along the $z$-axis. For
$v(y)$ along the $y$-axis, their results showed that the velocity
component $v(y)$ at different values of $y$ are highly correlated
and have a common phase across the cell's entire diameter. For
$v(z)$ along the $z$-axis, their results revealed that the
horizontal velocity component $v(z)$ at different height $z$
remain in phase along the $z$-axis mainly in the middle one-half
of the cell, while the horizontal velocity components $v(z)$ near
the upper and lower conducting plates gradually lag behind those
in the central region of the cell. Both of these results imply
that the motion of the central bulk fluid is spatially coherent
and the off-center oscillation is the dominant mode of the central
bulk fluid motion.

\begin{figure}
\center
\resizebox{1\columnwidth}{!}{%
  \includegraphics{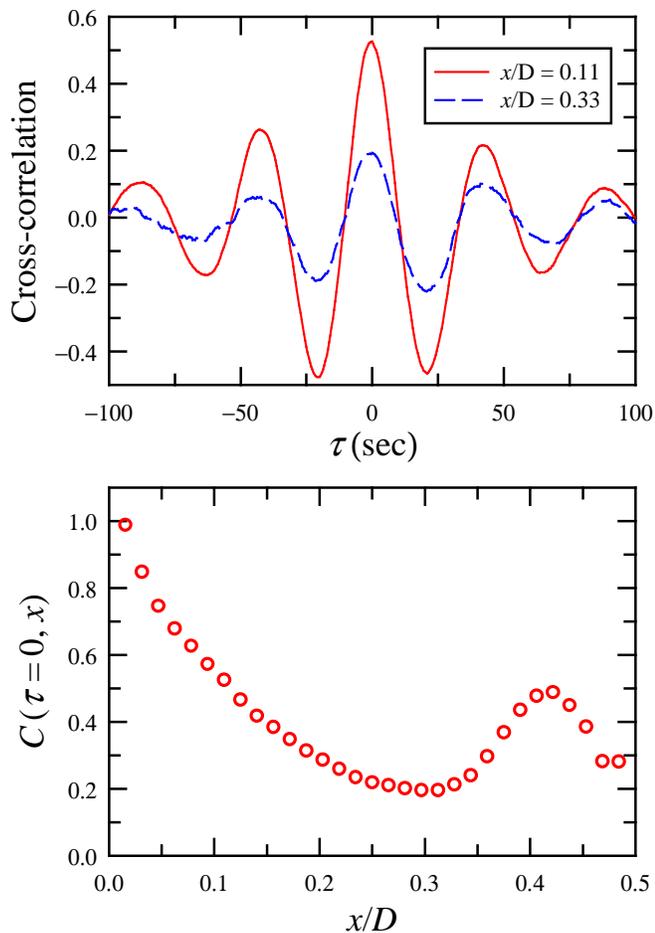}
}\caption{(a) The cross-correlation functions $C(\tau, x)$ between
$v(x,0)$ and $v(-x,0)$ for $x/D=0.11$ (solid line) and $0.33$
(dashed line). (b) The cross-correlation coefficient $C(\tau=0,
x)$ as a function of $x$.}\label{fig:fig13}
\end{figure}

The bulk off-center mode of the flow field can also be revealed by
investigating the variation along the $x$-axis of the $y$ velocity
component $v(x,0)$. Here we study the relationship between
$v(x,0)$ and $v(-x,0)$ for various values of $x(>0)$. If the
azimuthal rotation is the dominant motion of the LSC, $v(x,0)$
should anticorrelate with $v(-x,0)$, i.e., $v(x,0)$ is along the
$y$-direction when $v(-x,0)$ is along the $-y$-direction and vice
versa. While $v(x,0)$ correlates strongly with $v(-x,0)$ for the
situation that the off-center motion is the LSC's dominant mode,
i.e., $v(x,0)$ and $v(-x,0)$ are along the same direction. Figure
\ref{fig:fig12}(b) shows the typical time series of $v(x,0)$ and
$v(-x,0)$ for $x/D=0.11$. It is seen that $v(x,0)$ and $v(-x,0)$
are very similar to each other and along the same direction for
most of the time. In addition, both $v(x,0)$ and $v(-x,0)$ are
found to exhibit a well-defined periodic oscillation,
corresponding to the off-center oscillation of the LSC. To
characterize quantitatively these features, figure
\ref{fig:fig13}(a) shows the cross-correlation functions $C(\tau,
x)$ between $v(x,0)$ and $v(-x,0)$ for $x/D=0.11$ (solid line) and
0.33 (dashed line). It shows that both functions oscillate and
have a large positive peak located at $\tau=0$, indicating that
$v(x,0)$ correlates strongly with $v(-x,0)$. It is further found
that this peak exists for all $x$ and all horizontal velocity
oscillations of $v(x,0)$ have a common phase across the cell's
entire diameter. Figure \ref{fig:fig13}(b) shows the
cross-correlation coefficient $C(\tau=0, x)$ as a function of $x$.
One sees that $C(\tau=0, x)\gtrsim0.2$ for all $x$ and it has a
peak near the sidewall (around $x/D\simeq0.42$). The peak of
$C(\tau=0, x)$ near the sidewall represents the anticorrelation
between $\omega_{h,m}$ and $\omega_{c,m}$, i.e. the off-center
motion of the LSC, as illustrated in figures \ref{fig:fig7}(b) and
(c). Both the strong positive peak located at $\tau=0$ and the
oscillation of the cross-correlation functions $C(\tau,x)$ between
$v(x,0)$ and $v(x,0)$ for all $x$ reveal the fact that the
dominant motion of the LSC at the mid-height plane is indeed the
off-center oscillation, which provides another direct evidence for
the off-center mode of the LSC at the mid-height plane.

\section{Conclusion}

To conclude, we have made a detailed experimental study of the
oscillations of the large-scale circulation (LSC) in a cylindrical
turbulent Rayleigh-B\'{e}nard convection cell with aspect ratio
unity using water as working fluid. Direct spatial measurements of
both the temperature and velocity fields were carried out to study
the motion of the LSC.

Direct measurements of the horizontal velocity field at the cell's
mid-height plane using the PIV technique were made at
$Ra=3.0\times10^9$ and $6.0\times10^9$ and $Pr=4.3$. It is found
that the horizontal translational motion of the central bulk fluid
in the direction perpendicular to the vertical circulation plane
of the LSC is more probable, much stronger and more coherent than
the motions in all other directions. For the velocity component
$v(x,0)$ perpendicular to the LSC plane, $v(x,0)$ for all values
of $x$ are found to correlate strongly with each other, especially
for the velocities near the sidewall, and have a common phase
across the cell's entire diameter. Both of these results provide
direct evidences for the off-center mode of the LSC at the
mid-height plane.

The temperature measurement was performed over the Rayleigh-number
range $9\times10^8\leq Ra\leq6\times10^9$ and at fixed Prandtl
number $Pr=5.3$. At each $Ra$ the instantaneous azimuthal
temperature profile along the sidewall at three different heights,
i.e. $H/4$, $H/2$, and $3H/4$ from the bottom plate, were
measured. To study the motion of the LSC, we developed a new
method, the temperature-extremum-extraction (TEE) method. Using
this method, the azimuthal angular positions of the hot ascending
and cold descending flows at each height were obtained by making a
quadratic fit around the highest and lowest temperature readings,
respectively, and the line connecting these two positions is the
central line of the LSC band. The orientation of this line is thus
the orientation of the LSC and the distance between this line and
the cell's central vertical axis is defined as the off-center
distance. The motion of the LSC is therefore decomposed into two
different modes: the azimuthal mode and the translational or
off-center mode. When compared to the sinusoidal-fitting (SF)
method, it is found that, as far as the azimuthal motion of the
LSC is concerned, the two methods give the same information in
terms of both the orientation and the flow strength of the LSC,
while the TEE method can in addition determine the LSC's
off-center motion that is missed by the SF method. With the TEE
method, the LSC's central line is found to oscillate
time-periodically around the cell's central vertical axis with an
amplitude being nearly independent of the turbulent intensity,
leading to the off-center oscillation of the LSC.

Results obtained using the TEE method further reveal that both the
torsional and off-center modes of the LSC are manifestations of
the out-of-phase azimuthal positional oscillations of the hot
ascending and cold descending flows. When these two flows are near
the top and bottom plates, the resulting motion is torsional. When
they are near the mid-height plane, the resulting motion is the
off-center oscillation.

\begin{acknowledgments}

We gratefully acknowledge support of this work by the Hong Kong
Research Grants Council under Grant Nos. CUHK 403806 and 404307.

\end{acknowledgments}

\end{document}